\documentclass[twocolumn,aps,pra,superscriptaddress,showpacs]{revtex4-1}
\usepackage{graphicx}

\usepackage{subfigure}
\usepackage{epsfig}
\usepackage{dcolumn}   
\usepackage{bm}        
\usepackage{amssymb,amsmath,amsopn,amsfonts}
\usepackage{colordvi}
\usepackage{color}
\usepackage{multirow}
\usepackage{subeqnarray}
\usepackage{cases}

\usepackage{dsfont}

\newcommand{\ket}[1]{\ensuremath{\left| #1 \right\rangle}}

\newcommand{\ketbra}[2]{\ensuremath{\left| #1 \rangle\langle #2\right|}}

\newcommand{\Tr}{\operatorname{Tr}}
\def\eqref#1{\textup{(\ref{#1})}}

\begin{document}

\title{
Taking tomographic measurements for photonic qubits 88\,ns before they are created
}

\author{Zhibo Hou}
\affiliation{Key Laboratory of Quantum Information,University of Science and Technology of China, CAS, Hefei 230026, P. R. China}
\affiliation{Synergetic Innovation Center of Quantum Information and Quantum Physics, University of Science and Technology of China, Hefei 230026, P. R. China}
\author{Qi Yin}
\affiliation{Key Laboratory of Quantum Information,University of Science and Technology of China, CAS, Hefei 230026, P. R. China}
\affiliation{Synergetic Innovation Center of Quantum Information and Quantum Physics, University of Science and Technology of China, Hefei 230026, P. R. China}
\author{Chao Zhang}
\affiliation{Key Laboratory of Quantum Information,University of Science and Technology of China, CAS, Hefei 230026, P. R. China}
\affiliation{Synergetic Innovation Center of Quantum Information and Quantum Physics, University of Science and Technology of China, Hefei 230026, P. R. China}
\author{Han-Sen Zhong}
\affiliation{Key Laboratory of Quantum Information,University of Science and Technology of China, CAS, Hefei 230026, P. R. China}
\affiliation{Synergetic Innovation Center of Quantum Information and Quantum Physics, University of Science and Technology of China, Hefei 230026, P. R. China}

\author{Guo-Yong Xiang}
\email{gyxiang@ustc.edu.cn}
\affiliation{Key Laboratory of Quantum Information,University of Science and Technology of China, CAS, Hefei 230026, P. R. China}
\affiliation{Synergetic Innovation Center of Quantum Information and Quantum Physics, University of Science and Technology of China, Hefei 230026, P. R. China}
\author{Chuan-Feng Li}
\affiliation{Key Laboratory of Quantum Information,University of Science and Technology of China, CAS, Hefei 230026, P. R. China}
\affiliation{Synergetic Innovation Center of Quantum Information and Quantum Physics, University of Science and Technology of China, Hefei 230026, P. R. China}
\author{Guang-Can Guo}
\affiliation{Key Laboratory of Quantum Information,University of Science and Technology of China, CAS, Hefei 230026, P. R. China}
\affiliation{Synergetic Innovation Center of Quantum Information and Quantum Physics, University of Science and Technology of China, Hefei 230026, P. R. China}
\author{Geoff J. Pryde}
\affiliation{Centre for Quantum Computation and Communication Technology (CQC2T) and Centre for Quantum Dynamics, Griffith University, Brisbane, 4111, Australia}
\author{Anthony Laing}
\email{anthony.laing@bristol.ac.uk}
\affiliation{Quantum Engineering and Technology Laboratories, School of Physics and Department of Electrical and Electronic Engineering, University of Bristol, UK}

\begin{abstract}
We experimentally demonstrate that tomographic measurements can be performed for states of qubits before they are prepared.
A variant of the quantum teleportation protocol is used as a channel between two instants in time,
allowing measurements for polarisation states
of photons to be implemented 88~ns before they are created.
Measurement data taken at the early time and later unscrambled according to the results of the protocol's Bell measurements,
produces density matrices with an average fidelity of $0.90\pm0.01$ against the ideal states of photons created at the later time.
Process tomography of the time reverse quantum channel
finds an average process fidelity of $0.84\pm0.02$.
While our proof-of-principle implementation necessitates some post-selection,
the general protocol is deterministic and requires no post-selection to sift desired states and reject a larger ensemble.
\end{abstract}

\maketitle

\begin{figure*}[t]
  \centering
\center{\includegraphics[scale=.7]{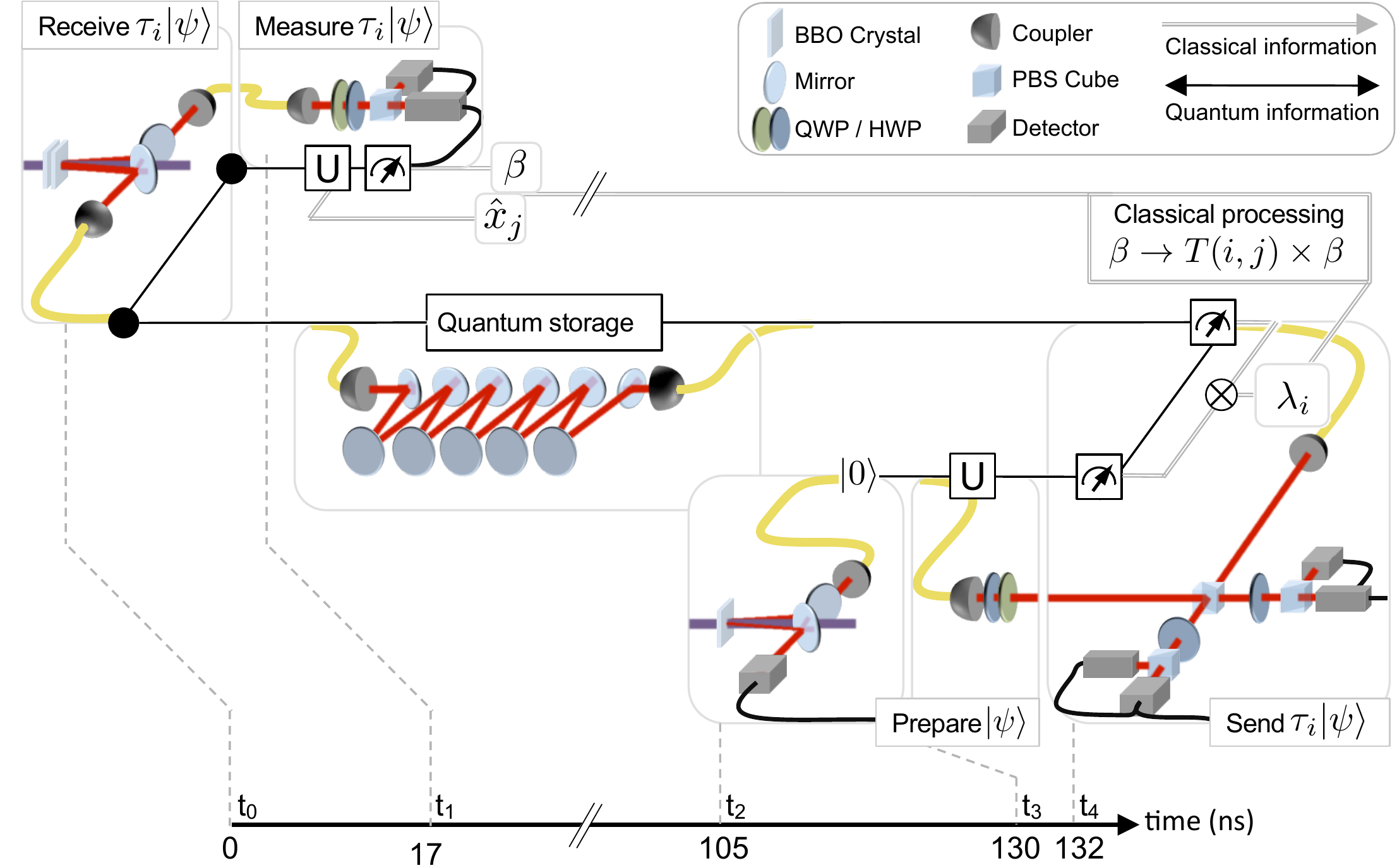}}
  \caption{\label{fig:timing}
Schematic for antedated quantum tomography protocol and experimental implementation.
($t_{0}$)
A pair of qubits in the entangled state $\ket{\Phi_{AB}^{+}}$ are encoded into the polarisation of the two-photon state that is produced from pumping a pair of crossed non-linear crystals.
($t_{1}$)
Qubit B from the entangled pair is measured along a direction $\hat{x}_{j}$, which we implement at $t=17$\,ns with a collection of waveplates, a polarising beamsplitter (PBS), and two single photon detectors.
Qubit A from this pair is stored in a quantum memory, realised here with a 31\,m storage line between a series of 11 mirrors.
($t_{2}$)
Qubit 3 is initialised: a single nonlinear crystal is pumped to generate a pair of photons at $t=105$\,ns, with one of the pair sent directly to a detector to herald the presence of the other.
($t_{3}$)
The state $\ket{\psi}$ is prepared on qubit 3 at $t=130$\,ns with a series of waveplates.
($t_{4}$)
Qubit A is retrieved from the quantum memory and measured together with qubit 3 in the Bell basis, which we execute at $t=132$\,ns with a PBS, a collection of waveplates, and four single photon detectors.
The Bell measurement effectively sends the state $\tau_{j} \ket{\psi}$ back through the quantum time channel to qubit B and on to the tomographic apparatus at $t_{1}$.
(See Appendix for further experimental details.)
   }
\end{figure*}

\begin{figure*}[t]
  \centering
\center{\includegraphics[scale=.7]{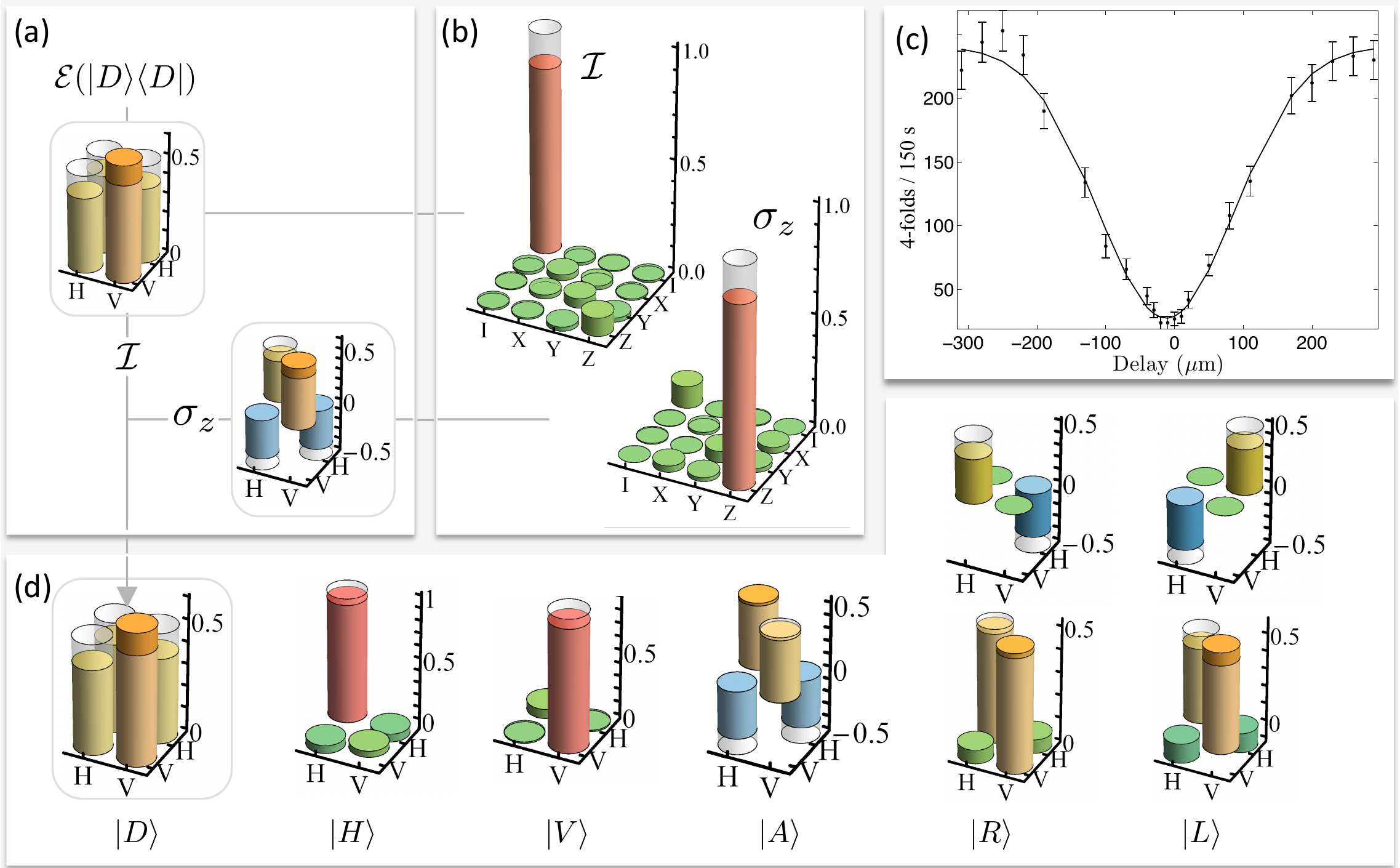}}
\caption{
Experimental results for antedated tomography.
(a)
With $\ket{D} \equiv \ket{+}$ prepared at time $t_{3}$ and sent at $t_{4}$ by a Bell measurement with result $\lambda_{0}$ or $\lambda_{3}$,
the ideal time channel to the tomography station at $t_{1}$ is the Identity or $\sigma_{z}$, respectively,
so that $\ket{D}$ or $\ket{A} \equiv \ket{-}$ is measured, respectively.
The real parts of the measured density matrices for both cases are displayed as solid colours with ideal values shown as the wire frame.
(b)
All six canonical polarisation states were prepared so process matrices for the time channel could be reconstructed,
the real parts of which are displayed.
(c)
The Bell measurement relies on Hong-Ou-Mandel interference between two photons generated at instants separated by 105\,ns.
The characteristic dip is shown for four-fold coincidence counts per 150\,s,
with a Gaussian fit (black line) to the experimental points
and error bars calculated from Poisson statistics.
(d)
For each of the six prepared states, data was post-processed according to our antedated tomography protocol.
Real parts of the reconstructed density matrices are respectively shown for prepared states $\ket{H}$, $\ket{V}$, $\ket{D}$, and $\ket{A}$;
for $\ket{R}$
and
$\ket{L}$,
the imaginary parts of the density matrices are shown in addition, below.
(See Appendix for numerical tables of density matrices and process matrices.)
}
\label{fgFringes}
\end{figure*}

Methods for characterising physical processes, found throughout science and engineering,
naturally assume that a system must first exist before measurements can be performed for it
\cite{moulder1995handbook, williams1996transmission, zewail2000femtochemistry}.
However, the transfer of quantum information between particles need not follow a chronology that is expected in classical physics
\cite{Lloy11quantum}.
According to quantum mechanics,
the state of a remote particle is instantaneously changed upon the measurement of a local particle with which it is entangled \cite{BrunCPS13},
while teleportation protocols predict that a state can sometimes be received before it is sent \cite{PereJMO00, Lloy11closed}.
A time symmetric picture of quantum mechanics explains these phenomena by viewing the act of a projective measurement as the preparation of a state that is sent backwards in time \cite{Aharonov2008, PhysRevA.79.052110, Oreshkov-NPhys-11-853}. 
But because the results of quantum measurements are random in general,
they cannot support superluminal communication or a communication channel with the past,
so conflicts with fundamental principles such as special relatively are avoided.
Any quantum state that is received before it is teleported will be mixed among a larger ensemble of transformed states from which it can be identified and sifted only at times after it is sent \cite{Svet11time}.
And the probability for an n-particle state to be faithfully received at a time before it is sent falls off exponentially.

These paradox-resolving clauses of time symmetry in quantum mechanics have
consigned its application to intrinsically post-selected experiments of philosophical interest
such as probabilistic simulations of closed time-like curves
\cite{Lloy11quantum, Lloy11closed}
or probabilistic demonstrations of nonlocality between particles before they interact via delayed choice experiments
\cite{PereJMO00, Jenn01experimental, MaNPhys-8-479, MegiPRL13}.
Here we experimentally demonstrate that time symmetry in quantum mechanics
can be used to perform measurements for quantum systems, ahead of time,
in a way that is not possible in classical physics.
Our protocol utilises all received states, including transformed states, and unscrambles the classically recorded measurement results after the time of sending,
to deterministically recover tomographic information \cite{Pari04quantum,Jame01measurement, ChuJMO-44-2455, PhysRevLett.78.390}.
Allowing state preparation to be performed after state measurement,
could open up new possibilities
for characterisation and verification \cite{JJW-NJP-2014, rbkNatComms2017}
in engineered quantum systems
\cite{PhysRevLett.74.4091, BarendsNatureSCQ, Carolan711}
and in quantum physics more generally
\cite{BlochNatPhysRev}.

Our antedated tomography protocol can be understood by first looking at the standard teleportation protocol
\cite{Benn93teleporting, Bouw97experimental, BoscBMH98,Marc03long,
Wang15quantum}.
A pair of particles encoded with qubits in an entangled state, such as
$\ket{\Phi_{AB}^{+}} = \frac{1}{\sqrt{2}} (\ket{0_{A}\,0_{B}} + \ket{1_{A}\,1_{B}})$,
are shared between between two agents, Alice and Bob.
Alice teleports the unknown state, $\ket{\psi}$, of a third particle to Bob,
by first performing an entangling (Bell) measurement between it and her particle from the entangled pair, then communicating the measurement result, $\lambda$,  to Bob through a classical channel.
The initial 3-particle state, $\ket{\Phi_{AB}^{+}}\ket{\psi_{3}}$, written in the basis of Bell states for particles $A$ and 3,
\begin{align}
\ket{\Phi_{AB}^{+}}\ket{\psi_{3}} = \;
    ( \;
    & \ket{\Phi_{A3}^{+}} \tau_{0} \ket{\psi_{B}} 
+  \ket{\Psi_{A3}^{+}} \tau_{1} \ket{\psi_{B}} 
\\
 + \;
 & \ket{\Psi_{A3}^{-}} \tau_{2} \ket{\psi_{B}} 
+  \ket{\Phi_{A3}^{-}} \tau_{3} \ket{\psi_{B}} 
\; )/2, \nonumber
\label{eqtele01}
\end{align}
shows that the Bell measurement projects Bob's particle into a state, $\tau\ket{\psi}$, that is equal to $\ket{\psi}$ up to one of four unitary transformations.
The transformation
occurs randomly from a set that is related simply to the Identity and Pauli operators,
$\tau \in \{\tau_{i}\}^{3}_{i=0} = \{\mathcal{I}, \sigma_{1},i\sigma_{2},\sigma_{3}\}$,
but
is correlated with the results of the Bell measurement, $\lambda \in \{\lambda_{i}\}^{3}_{i=0}$.
When Bob learns the value of $\lambda$ from Alice, he can correct for the corresponding $\tau$ on his particle and set its state equal to $\ket{\psi}$.

If Bob measures his particle before Alice performs her Bell measurement or before he learns the value of $\lambda$,
he forgoes the opportunity to undo the $\tau$ transformation.
For approximately one-quarter of experimental trials, the Bell measurement outcome will equal $\lambda_{0}$,
which indicates that $\tau$ is equal to the Identity, $\tau_{0}\ket{\psi}=\ket{\psi}$.
These are cases where Bob's measurement results are consistent with $\ket{\psi}$ when he applies no operation to his particle.
Here, Bob effectively possesses $\ket{\psi}$ before it is teleported, or even before it exists,
but he must wait for the classical signal from Alice before he can identify these cases within the full ensemble.
In the other three quarters of experimental trials, the classical message, $\{\lambda_{i}\}^{3}_{i=1}$, indicates to Bob
which of the other transformations he should have undone to set his particle equal to $\ket{\psi}$.
In general for these cases, when Bob measures first, his results are not consistent with\,$\ket{\psi}$.

While the $\tau$ transformations prevent a communication channel to the past,
they need not be an obstacle to the characterisation of $\ket{\psi}$,
and in some scenarios are beneficial.
A particular measurement operator, $\vec{n} \cdot \vec{\sigma}$,
set for a projective measurement on Bob's particle
is randomly transformed to implement a different basis,
$\vec{b} \cdot \vec{\sigma} = \tau_{i}  (\vec{n} \cdot \vec{\sigma})  \tau^{\dagger}_{i}$,
for measuring $\ket{\psi}$
(where $\vec{n}$ and $\vec{b}$ are unit vectors, and $\vec{\sigma}$ is the Pauli vector).
Bob learns the new setting, $\vec{b}$, when he learns the result, $\lambda_{i}$, of the Bell measurement.
Vector components transform such that
$b_{j} = n_{j}$
for $i=j$ or $i=0$, whereas $b_{j} = -n_{j}$ otherwise.
One characterisation strategy results from the observation that any fixed measurement setting,
$\vec{n}$,
with non-zero coefficients along each direction,
$\hat{x}_{1}, \hat{x}_{2}, \hat{x}_{3}$,
will lead to an informationally complete basis set,
$\{\tau_{i} (\vec{n} \cdot \vec{\sigma}) \tau^{\dagger}_{i}\}^{3}_{i=0}$
over an ensemble of trials.
Such a basis set is more stable when
$\vec{n}$ is not too unequally weighted across $\hat{x}_{1}, \hat{x}_{2}, \hat{x}_{3}$.

Our experiments implement a different strategy
to optimally reduce the effect of statistical errors
by realising measurements in exactly the Pauli bases
 \cite{WOOTTERS1989363}.
Setting $\vec{n}$
along exactly one of the directions $\hat{x}_{j}$
means that the corresponding $\sigma_{j}$ intended to measure Bob's particle,
leads to a measurement on
$\ket{\psi}$
in the same basis,
but with the eigenvalues sometimes reversed,
as described above.
Our antedated tomography protocol therefore requires Bob to
perform a complete set of tomographic measurements in the Pauli bases on his particle,
and unscramble the classical data
to recover measurement results for $\ket{\psi}$
when he later learns the $\lambda$ values.

The protocol and its resources are as follows:
Alice and Bob share an ensemble of Bell states and a classical communication channel.
On his qubits from the entangled pairs,
Bob performs a compete set of tomographic measurements in the Pauli bases
and records his measurement choices (two bits each) with respective results (one bit each) in a classical memory.
Alice stores her qubits from the entangled pairs in a quantum memory.
When Alice acquires an ensemble of qubits for which she would like Bob's existing tomographic data to apply,
she performs Bell measurements between these and her qubits retrieved from the quantum memory.
Alice classically transmits the Bell measurement results (two bits each) to Bob,
which he marries with his stored classical data
and performs post-processing of expectation values.
Bob's $k$th measurement result,
$\beta_{k}\in\{-1,+1\}$,
is transformed,
$\beta_{k} \rightarrow T(i_{k},j_{k}) \times \beta_{k}$,
according to his $k$th Pauli basis choice
$\sigma^{(k)}_{j}$
and Alice's $k$th Bell measurement result,
$\lambda^{(k)}_{i}$.
$T$ leaves $\beta_{k}$ unchanged when $i_{k}=j_{k}$ or when $i_{k}=0$;
$T$ flips the sign on $\beta_{k}$ otherwise.

We implemented our antedated tomography protocol on ensembles of photons, as shown in Fig. 1.
At time $t=0$ ($t_{0}$)
a laser pulse pumps a pair of crossed non-linear crystals
to produce a pair of photons that are polarisation-entangled \cite{PhysRevA.60.R773}
in the state $\ket{\Phi_{AB}^{+}}=\frac{1}{\sqrt{2}}(\ket{H_{A},H_{B}}+\ket{V_{A},V_{B}})$,
where a horizontally polarised photon encodes $\ket{0}$ and a vertically polarised photon encodes $\ket{1}$.
At $t=17$\,ns ($t_{1}$) one photonic qubit from this pair (photon B) is measured in a Pauli basis,
with the basis choice and measurement result recorded on a classical computer.
The other photon from this pair (photon A) is directed to a $31$\,m storage channel that serves as a quantum memory.
At $t=105$\,ns ($t_{2}$) a second pair of (polarisation-unentangled) photons is created,
one of which is immediately directed to a detector to herald the creation of the other, photon 3.
At $t=130$\,ns ($t_{3}$) photon 3 is prepared in a desired state $\ket{\psi_{3}}$.
Finally, at $t=132$\,ns ($t_{4}$), photon 3 and the delayed photon A are measured in the Bell state analyser,
by impinging both photons simultaneously onto the two input ports of a polarising beamsplitter followed by waveplates \cite{MegiPRL13}.
The measurement result is recorded with the classical computer.
(See Appendix for further experimental details.)

The apparatus at $t_{3}$ was configured to prepare
the six canonical polarisation states, $\{ H,V,D,A,R,L \}$, on photon 3,
which respectively represent
$\ket{\psi_{3}} \in
\{
\ket{0}, \ket{1}, \ket{\pm}=\frac{1}{\sqrt{2}}(\ket{0} \pm \ket{1}), \ket{\pm_{i}}=\frac{1}{\sqrt{2}}(\ket{0} \pm i \ket{1})
\}$.
For ensembles of many photons prepared in each of these states,
the apparatus at $t_{1}$ measured expectation values in the three Pauli bases.
The Bell state analyser at $t_{4}$ in our implementation,
was able to deterministically resolve two of the four Bell states,
which were chosen as $\ket{\Phi^{\pm}}$,
corresponding to eigenvalues $\lambda_{0}$ and $\lambda_{3}$ respectively.
Data was post-processed
such that all expectation values were unaltered 
when $\lambda_{0}$ was found on photons A and 3,
whereas the sign was flipped on expectation values measured for $\sigma_{1}$ and $\sigma_{2}$
when $\lambda_{3}$ was found.
Density matrices, as shown in Fig. 2(d),
calculated via maximum likelihood estimation \cite{Pari04quantum,Jame01measurement},
were found to have an average
fidelity of $0.90\pm0.01$
with ideally prepared states, $\ket{\psi_{3}}$
(See Appendix for tables of individual density matrices.)

Using the set of prepared $\ket{\psi_{3}}$
together with measured density matrices
(shown in Fig. 2(a) for $\ket{\psi_{3}} = \ket{D}$)
from data that were unaltered but sorted according to $\lambda_{0}$ and $\lambda_{3}$,
the quantum time-channel
connecting state preparation at $t_{3}$ and state measurement at $t_{1}$
was characterised using full quantum process tomography
\cite{ChuJMO-44-2455, PhysRevLett.78.390}.
Experimentally, the time-channel consists of
an entangled state,
a quantum memory in the form of a $31$\,m photonic delay line,
and a Bell state analyser.
In the cases where $\lambda_{0}$ was found in the Bell state analyser,
the time-channel should ideally be the Identity operation,
while when $\lambda_{3}$ is measured,
the time-channel should ideally be the $\sigma_{3}$ operation.
Both process matrices were reconstructed, as shown in Fig. 2(b),
with fidelities found to be $0.84 \pm 0.02$ and $0.83 \pm 0.02$ for the Identity and $\sigma_{3}$ respectively.
(See Appendix for further details of process tomography.)

The quality of quantum interference \cite{HongOM87} between pairs of photons at the beamsplitter in the Bell state analyser, is critical to its correct operation.
With one photon in the pair having passed through a 31\,m free-space channel,
a quantum interference visibility of 0.89$\pm$0.01 was measured, as shown in Fig 2(c).

Much of the history of quantum science
is the evolution of philosophically contentious concepts
to practical applications.
While causal order
and time symmetry in quantum mechanics are fundamentally interesting
\cite{PhysRevA.79.052110, Oreshkov:2012aa}, 
removing the constraint of state-preparation before state-measurement
could open up new avenues for existing characterisation methods
\cite{JJW-NJP-2014, rbkNatComms2017}.
If entangling measurements between different types of particles,
such as atomic and molecular ions \cite{Nat-Wolf-530-457},
become feasible, then a tomographic capability that is more readily available
at an earlier time or for a certain atomic species \cite{Nat-Bal-528-384},
can be deployed to characterise states encoded in less accessible physical systems at later times.
In solid state quantum processors \cite{NatPhot-Gao-9-363},
tomography could be performed ahead of time using photons
for subsequently prepared states of spins,
or for the mediating electronic spin,
with the long lived nuclear spin serving as a quantum memory.
Because the general antedated quantum tomography protocol is deterministic,
it could prove useful for characterising or verifying larger physical systems
that are efficiently described by matrix product states \cite{Cram10efficient}.
\\
We are grateful to S. Popescu and T. Short for helpful discussions.
This work was supported by the
National Natural Science Foundation of China (Grant No. 11574291, 61222504) and National Key R \& D Program (Grant No. 2016YFA0301700),
the US Army Research Office (ARO) Grant No. W911NF-14-1-0133, and
the Australian Research Council (DP140100648).
AL acknowledges fellowship support from the Engineering and Physical Sciences Research Council (EPSRC, UK).

\clearpage
\onecolumngrid
\section{Appendix}

\noindent The experimental apparatus was comprised of five sections:\\
\indent 1. Entangled state generation,\\
\indent 2. Tomographic setup for photon B,\\
\indent 3. Delay line for photon A,\\
\indent 4. Generation and state preparation for photon 3, and\\
\indent 5. Bell state measurement.\\
Results and analyses are presented for:\\
\indent 6. Experimental results for single qubit tomography, and\\
\indent 7. Experimental results for process tomography.
 
\subsection{1. Entangled state generation}
Pulses from a mode-locked Ti-sapphire laser with a duration of 140\,fs, a repetition rate of 76\,MHz, and a central wavelength of 780\,nm first passed through a frequency doubler.
The resultant ultraviolet pulses with a power of 200\,mW then (at t=0\,ns) pumped a pair of sandwiched beam-like-type cut BBO crystals, each of thickness 1\,mm,
to produce a stream of photon pairs, entangled in their polarisation.
Temporal and spatial compensation (TSC) was made with LiNbO$_3$ and YVO$_4$ crystals \cite{Zhan15experimental}.
The photonic qubits were then distributed to Alice and Bob via a pair of 3\,m single mode fibres (SMF).
Half waveplates (HWP) and tilted quarter waveplates (QWP) at both ports of the fibres were used to maintain the polarisation reference frame.

With the photonic qubits on Alice's side having passed through the 31\,m delay line and a 2\,m SMF,
both qubits were measured in the Pauli basis (details below),
and their density matrix reconstructed by the maximum likelihood estimation algorithm \cite{Pari04quantum}.
The quantum state that generates the experimental data with highest probability is
\begin{equation}
\rho_\text{MLE}=\left(
         \begin{array}{cccc}
           0.486 & 0.026 + 0.007i & -0.031 - 0.009i & 0.446 + 0.112i \\
           0.026 - 0.007i & 0.018 & -0.001 + 0.015i & 0.035 + 0.014i \\
            -0.031 + 0.009i & -0.001 - 0.015i & 0.021 & -0.020 - 0.017i \\
           0.446 - 0.112i & 0.035 - 0.014i & -0.020 + 0.017i & 0.475 \\
         \end{array}
       \right),
\end{equation}
which has a fidelity with the ideal state of $0.927\pm0.001$, where the error is the standard deviation from 100 numerical simulations with Poissonian statistics.

\subsection{2. Tomographic setup for photon B}
From each pair, the photon on Bob's side, photon B, was immediately injected into its tomographic set up and measured at t=17\,ns,
due to the overall optical path length from photon generation, including the approx. 0.6\,m in this section.
The module contained a HWP, a QWP, a polarisation beam splitter (PBS), and a single photon detector (SPD), to realise arbitrary projective measurements on qubits encoded in the polarisation of photons.
This photon was finally passed through an interference filter (FWHM of 8\,nm ) before detection.
Waveplate settings were chosen to implement measurements in the three Pauli bases,
and results were stored with respect to these settings.

\subsection{3. Delay line for photon A}
The photon on Alice's side, photon A, was directed to a delay line, comprised of eleven mirrors which reflected the photon through 31\,m of free-space before coupling to 2\,m of SMF.
A telescope was implemented with three concave lenses, resulting in an overall collection efficiency of approx. 50\% including losses from reflection and fibre insertion.

\subsection{4. Generation and sate preparation for photon 3}
After 8 consecutive pump pulses (105 ns), a second beam-like BBO crystal was pumped to generate pairs of (unentangled) photons,
with one photon from each pair directly detected as a trigger, after passing through an interference filter (FWHM of 8\,nm).
The other photon from this pair, photon 3, passed through a 5\,m SMF, after which its state was prepared by a combination of HWP and QWP.

\subsection{5. Bell state measurement}
Photon 3 and photon A impinged simultaneously on the first PBS.
Two HWPs, one at each PBS exit port, had optics axes set at 22.5$^\circ$, while a tilted QWP ensured the correct phase was implemented.
This configuration \cite{MegiPRL13}, deterministically resolves two of the four Bell states, such that even ($+ + , - -$) or odd ($+ - , - +$) measurement statistics are observed for states 
$\ketbra{\Phi^+}{\Phi^+}$
and
$\ketbra{\Phi^-}{\Phi^-}$
respectively.
The SMF and interference filters (FWHM of 2 nm) prior to detection reduced spatial and temporal difference between the two photons to produce a Hong-Ou-Mandel \cite{HongOM87} interference visibility of 0.89$\pm$0.01 at a pumping power of 200 mW.

\subsection{6. Experimental results for single qubit tomography}
Tomographic data for photon B
were collected over 150\,s at a pumping power of 200 mw
for each of the six states prepared on photon 3 and for each Pauli basis.
Data were grouped into two sets based on the result of the associated Bell state measurement,
$\ketbra{\Phi^\pm}{\Phi^\pm}$,
and respective single qubit density matrices, $\rho_{\Phi^{+}}$ and $\rho_{\Phi^{-}}$, were calculated with maximum likelihood estimation.
Table~\ref{table:states} shows these, while Table~\ref{table:statesC} shows density matrices from the combined data sets as described in the main text. Errors on fidelities are the standard deviations from 100 simulated results with Poissonian statistics. The average fidelity over all six states and two Bell state measurements is $0.90\pm0.01$.
\begin{table}[h]
  \caption{\label{table:states}
  Reconstructed single qubit density matrices $\rho_{\Phi^{+}}$ and $\rho_{\Phi^{-}}$ for all six states $\ket{\phi_{3}}$ prepared on photon 3,
  alongside their corresponding fidelities with errors.}
  \begin{tabular}{c c c c c }
    \hline
    $\ket{\phi}$ & $\rho_{\Phi^+}$ & $F(\rho_{\Phi^+},\ket{\phi})$ & $\rho_{\Phi^-}$ & $F(\sigma_{3}\rho_{\Phi^-}\sigma_{3},\ket{\phi})$\\
    \hline
    $\ket{H}$ &        $\left(
                       \begin{array}{cc}
                         0.94 & -0.02 + 0.06i \\
                         -0.02 - 0.06i & 0.06 \\
                       \end{array}
                     \right)      $          & 0.94$\pm0.03$
                     &        $\left(
                       \begin{array}{cc}
                         0.96 & 0.11 - 0.02i \\
                         0.11 + 0.02i & 0.04 \\
                       \end{array}
                     \right)      $          & 0.96$\pm0.03$\\
                     
    $\ket{V}$ &        $\left(
                       \begin{array}{cc}
                         0.06& -0.13 - 0.09i \\
                         -0.13 + 0.09i & 0.94 \\
                       \end{array}
                     \right)      $          & 0.94$\pm0.02$
                     &        $\left(
                       \begin{array}{cc}
                         0.08& -0.11-0.00i \\
                         -0.11+0.00i & 0.92 \\
                       \end{array}
                     \right)      $          & 0.92$\pm0.02$\\
    $\ket{D}$ &        $\left(
                       \begin{array}{cc}
                         0.40 & 0.38 + 0.01i \\
                         0.38 - 0.01i & 0.60 \\
                       \end{array}
                     \right)      $          & 0.88$\pm0.03$
                     &        $\left(
                       \begin{array}{cc}
                         0.40 & -0.39 - 0.11i \\
                         -0.39 + 0.11i & 0.60 \\
                       \end{array}
                     \right)      $          & 0.89$\pm0.03$\\
    $\ket{A}$ &        $\left(
                       \begin{array}{cc}
                         0.53 & -0.37 - 0.02i \\
                         -0.37 + 0.02i & 0.47 \\
                       \end{array}
                     \right)      $          & 0.87$\pm0.03$
                     &        $\left(
                       \begin{array}{cc}
                         0.52 & 0.38 + 0.12i \\
                         0.38 - 0.12i & 0.48 \\
                       \end{array}
                     \right)      $          & 0.88$\pm0.03$\\
    $\ket{R}$ &        $\left(
                       \begin{array}{cc}
                          0.46 & -0.01 - 0.40i \\
                         -0.01 + 0.40i & 0.54 \\
                       \end{array}
                     \right)      $          & 0.90$\pm0.03$
                     &        $\left(
                       \begin{array}{cc}
                          0.50 & -0.14 + 0.33i \\
                         -0.14 - 0.33i & 0.50 \\
                       \end{array}
                     \right)      $          & 0.83$\pm0.02$\\
    $\ket{L}$ &        $\left(
                       \begin{array}{cc}
                         0.40 & -0.12 + 0.38i \\
                         -0.12 - 0.38i & 0.60 \\
                       \end{array}
                     \right)      $          & 0.88$\pm0.04$
                     &        $\left(
                       \begin{array}{cc}
                         0.45 & 0.10 - 0.37i \\
                         0.10 + 0.37i & 0.55 \\
                       \end{array}
                     \right)      $          & 0.87$\pm0.03$\\
\hline
  \end{tabular}
  \end{table}
    \begin{table}[h]
 \centering{
  \caption{\label{table:statesC}
  Reconstructed single qubit density matrices $\rho_3$ for all six states $\ket{\phi_{3}}$ prepared on photon 3,
  alongside their corresponding fidelities with errors.}
  \begin{tabular}{c c c }
    \hline
    $\ket{\phi_3}$ & $\rho_3$ & $F(\rho_3,\ket{\phi_3})$\\
    \hline
    $\ket{H}$ &        $\left(
                       \begin{array}{cc}
                         0.95 & -0.06 + 0.04i \\
                         -0.06 - 0.04i & 0.05 \\
                       \end{array}
                     \right)      $          & 0.95$\pm0.02$\\
    $\ket{V}$ &        $\left(
                       \begin{array}{cc}
                         0.07& -0.01 - 0.05i\\
                         -0.01+ 0.05i & 0.93 \\
                       \end{array}
                     \right)      $          & 0.93$\pm0.02$\\
    $\ket{D}$ &        $\left(
                       \begin{array}{cc}
                         0.40 & 0.39 + 0.06i \\
                         -0.39 - 0.06i & 0.60 \\
                       \end{array}
                     \right)      $          & 0.89$\pm0.02$\\
    $\ket{A}$ &        $\left(
                       \begin{array}{cc}
                         0.52 & -0.37 - 0.07i \\
                         -0.37 + 0.07i & 0.48 \\
                       \end{array}
                     \right)      $          & 0.87$\pm0.02$\\
    $\ket{R}$ &        $\left(
                       \begin{array}{cc}
                          0.48 & 0.06 - 0.37i \\
                         0.06 + 0.37i & 0.52 \\
                       \end{array}
                     \right)      $          & 0.87$\pm0.02$\\
    $\ket{L}$ &        $\left(
                       \begin{array}{cc}
                         0.43 & -0.11 + 0.38i \\
                         -0.11 - 0.38i & 0.57 \\
                       \end{array}
                     \right)      $          & 0.88$\pm0.02$\\
    \hline
  \end{tabular}}
  \end{table}

\subsection{7. Experimental results for process tomography}
Following \cite{Jeif03quantum}, $S$ is a positive semidefinite operator on a Hilbert space $\mathcal{H}\otimes\mathcal{K}$
that maps the quantum state $\rho_\text{in}$ on a Hilbert space $\mathcal{H}$
to the output state $\rho_\text{out}$ on a Hilbert space $\mathcal{K}$,
\begin{equation}\label{EQ: process tomo}
  \rho_\text{out}=\Tr_H[S\rho_\text{in}^T\otimes1_\mathcal{K}],
\end{equation}
where $1_\mathcal{K}$ is an identity operator on $\mathcal{K}$, and $T$ denotes the matrix transposition.
$S_\text{MLE}$ is found from an iterative method to generate the experimental data with maximum probability.
As described in the main text, two different processes $S^\pm$ are respectively associated with
the two Bell measurement results $\ketbra{\Phi^\pm}{\Phi^\pm}$.
The ideal and maximum likelihood estimators of $S^\pm$ are obtained via the algorithm in \cite{Jeif03quantum}
using the experimental data that produced the density matrices in Table \ref{table:states}.

The $S^\pm$ matrices are simply transformed into $\chi^\pm$
from the standard definition \cite{Obri04quantum}
\begin{equation}\label{Eq:process tomo other}
  \rho_\text{out}=\sum_{m,n=0}^{d^2-1}\chi_{mn}\sigma_m\rho_\text{in}\sigma_n
\end{equation}
with Identity and Pauli operators $\sigma_{m}$,
such that
\begin{equation}\label{Eq:S chi connection2}
  \chi=U_2^\dagger U_1^\dagger \, S \,\, U_1U_2
\end{equation}
with
\begin{equation}
  U_1=\left(
          \begin{array}{cccc}
            1 & 0 & 0 & 0 \\
            0 & 0 & 1 & 0 \\
            0 & 1 & 0 & 0 \\
            0 & 0 & 0 & 1 \\
          \end{array}
        \right),
        \;\;\;\;\;\;\;\;
 U_2=\frac{1}{2}\left(
    \begin{array}{cccc}
      1 & 0 & 0 & 1 \\
      0 & 1 & -i & 0 \\
      0 & 1 & i & 0 \\
      1 & 0 & 0 & -1 \\
    \end{array}
  \right)
\end{equation}
to give
\begin{equation}\label{Eq:chi+ MLE}
  \begin{aligned}
      \chi^+_\text{ideal}=\left(
          \begin{array}{cccc}
            1& 0 & 0 & 0 \\
            0 & 0 & 0 & 0 \\
            0 & 0& 0 & 0 \\
            0 & 0 & 0 & 0 \\
          \end{array}
        \right),\quad
    \chi^+_\text{MLE}
    =\left(
          \begin{array}{cccc}
            0.84 & -0.01+0.06i & 0.00+0.06i & -0.01-0.03i \\
            -0.01-0.06i & 0.03 & 0.02+0.01i & -0.01+0.00i \\
            0.00-0.06i & 0.02-0.01i & 0.04 & -0.02+0.01i \\
            -0.01+0.03i & -0.01-0.00i & -0.02-0.01i & 0.09 \\
          \end{array}
        \right),
  \end{aligned}
\end{equation}
and
\begin{equation}\label{Eq:chi-MLE}
  \begin{aligned}
      \chi^-_\text{ideal}=\left(
          \begin{array}{cccc}
            0& 0 & 0 & 0 \\
            0 & 0 & 0 & 0 \\
            0 & 0& 0 & 0 \\
            0 & 0 & 0 & 1 \\
          \end{array}
        \right),\quad
    \chi^-_\text{MLE}
    =\left(
          \begin{array}{cccc}
            0.10 & -0.00+0.01i & 0.01+0.07i & 0.00+0.12i \\
            -0.00-0.01i & 0.01 & 0.00-0.00i & 0.03+0.01i \\
            0.01-0.07i & 0.00+0.00i & 0.05 & 0.02+0.00i \\
            0.00-0.12i & 0.03-0.01i & 0.02-0.00i & 0.83 \\
          \end{array}
        \right).
  \end{aligned}
\end{equation}

The process fidelities are immediately available in this representation as
$0.84 \pm 0.02$ and $0.83 \pm 0.02$ for the Identity and $\sigma_{3}$ respectively
where the errors on the fidelities are the standard deviations from 100 simulated results with Poissonian statistics.

\end{document}